\documentclass[aps,prb,10pt,twocolumn,amsmath,amssymb,citeautoscript,
               superscriptaddress, longbibliography]{revtex4-2}
\usepackage[utf8]{inputenc}
\usepackage{graphicx}
\usepackage{dcolumn}
\usepackage{array}
\usepackage{amssymb}
\usepackage{amsmath}
\usepackage{float}
\usepackage{multirow}
\usepackage{braket}
\usepackage[normalem]{ulem}
\usepackage{bm}
\renewcommand{\vec}[1]{\mathbf{#1}}
\usepackage[hang]{footmisc}
\usepackage{hyperref}
\hypersetup{pdfnewwindow=true, colorlinks=true, linkcolor=blue, anchorcolor=blue, citecolor=blue, filecolor=blue, menucolor=blue, urlcolor=blue}

\newcommand{\cri}{CrI$_3$} 

\begin{document}
\title{Frequency splitting of chiral phonons from broken time reversal symmetry in CrI$_3$}
\author{John Bonini}
\affiliation{Center for Computational Quantum Physics, Flatiron Institute, 162 5th Avenue, New York, New York 10010,
USA}
\author{Shang Ren}
\affiliation{Department of Physics and Astronomy, Rutgers University, Piscataway, New Jersey 08845-0849, USA}
\author{David Vanderbilt}
\affiliation{Department of Physics and Astronomy, Rutgers University, Piscataway, New Jersey 08845-0849, USA}
\author{Massimiliano Stengel}
\affiliation{Institut de Ci\`{e}ncia de Materials de Barcelona (ICMAB-CSIC), Campus UAB, 08193 Bellaterra, Spain}
\affiliation{ICREA-Instituci\'{o} Catalana de Recerca i Estudis Avan\c{c}ats, 08010 Barcelona, Spain}
\author{Cyrus E. Dreyer}
\affiliation{Department of Physics and Astronomy, Stony Brook University, Stony Brook, New York, 11794-3800,
  USA}
\affiliation{Center for Computational Quantum Physics, Flatiron Institute, 162 5th Avenue, New York, New York 10010,
USA}
\author{Sinisa Coh}
\affiliation{Materials Science and Mechanical Engineering, University of California Riverside, CA 92521, USA}

\date{\today}
\begin{abstract}
Conventional approaches for lattice dynamics based on static interatomic forces do not fully account for the effects of time-reversal-symmetry breaking in magnetic systems. Recent approaches to rectify this involve incorporating the first-order change in forces with atomic velocities under the assumption of adiabatic separation of electronic and nuclear degrees of freedom. In this work, we develop a first-principles method to calculate this velocity-force coupling in extended solids, and show via the example of ferromagnetic CrI$_3$ that, due to the slow dynamics of the spins in the system, the assumption of adiabatic separation can result in large errors for splittings of zone-center chiral modes. We demonstrate that an accurate description of the lattice dynamics requires treating 
magnons and phonons
on the same footing.
\end{abstract}
\maketitle

The atomic vibrations that are present in molecules and solids at zero and
finite temperature play a crucial role in their thermodynamic and transport
properties. First-principles calculations based on
density-functional theory (DFT) have been established as a powerful tool for
understanding and predicting lattice-dynamical properties, including phonon dispersion \cite{gonze1997,Baroni_2001}
and electron-phonon coupling
\cite{Giustino2017,Monserrat2018}. The key quantity
underlying the calculation of lattice dynamics is the interatomic force constant
(IFC) matrix, which is constructed by finding derivatives of the nuclear
forces with respect to nuclear positions, either directly via finite
displacements
or though density functional perturbation
theory \cite{gonze1997,Baroni_2001}.

In presence of magnetic ordering,
the change in the electronic ground state 
compared to the nonmagnetic case propagates to the IFCs~\cite{PhysRevB.65.184422,Lee2011,Hong2012,Wang2021,Wu2022}.
However, since it is defined and calculated as a static response function, the IFC matrix is invariant under time reversal by construction.
Thus, a description of the nuclear dynamics based solely on the IFCs will not correctly reflect the vibration mode degeneracies in a magnetic system; instead, the phonon frequency spectrum will be determined by
the \emph{nonmagnetic} symmetry group.

There has been significant recent work on the explicit inclusion of time-reversal symmetry
(TRS) breaking in the nuclear equations of motion via the
\emph{velocity} dependence of the interatomic forces, applied to models
\cite{qin12_berry_curvat_phonon_hall_effec,
  PhysRevLett.123.255901,PhysRevB.105.064303,
  mead79_deter_born_oppen_nuclear_motion} and magnetic molecules \cite{Bistoni2021}. This
``velocity-force'' coupling can be obtained from the nuclear Berry curvature,
which describes the evolution of the phase of the electronic wavefunction with
changes in nuclear coordinates
\cite{qin12_berry_curvat_phonon_hall_effec,PhysRevLett.123.255901,Bistoni2021}.
A key result of including this coupling is that degenerate vibrational modes may
split into nondegenerate chiral modes \cite{Zhang2015} with a well-defined
finite angular momentum \cite{coh19_class_mater_with_phonon_angul,Bistoni2021},
even at the Brillouin-zone center. Thus, the correct treatment of magnetic symmetry is
crucial for elucidating the role of atomic vibrations in thermal Hall
\cite{Grissonnanche2019,PhysRevLett.123.255901,Chen2020,Li2020,PhysRevLett.127.247202}
and other effects involving TRS-broken lattice dynamics
\cite{Zhang2014,Zhu2018,Chen2018,Mentink2019,Yin2021,Chen2022,Baydin2022,Juraschek2022}.

A key assumption underlying the velocity-force approach in previous works
\cite{qin12_berry_curvat_phonon_hall_effec,PhysRevLett.123.255901,PhysRevB.105.064303,mead79_deter_born_oppen_nuclear_motion,Bistoni2021}
is that the time scale for electronic dynamics is fast compared to nuclear
dynamics. However, this  may completely break down in some systems, e.g., when
the nuclear Berry curvature results from nuclei coupling to \emph{spins}, whose
dynamics are not necessarily faster than the atomic vibrations.
The breakdown of the adiabatic picture could have a
profound effect on
the predicted splitting of chiral phonon modes.

In this work, we illustrate such a situation using the bulk-layered magnetic
insulator \cri{} as an example system.
We first develop a DFT methodology amenable to both molecules and solids
for computing phonons in the presence of velocity-force coupling.
We apply this method
to calculate the zone-center
phonons in \cri{}, demonstrating the splitting
of otherwise degenerate phonons into chiral modes. Next, we show that the velocity-force
response is dominated by the canting of the spins on the Cr sites caused by
atomic displacements. The dynamics of such spin canting are characterized by the
zone-center magnon frequencies. The fact that magnon frequencies in \cri{} are
on the same order or smaller than the optical phonon frequencies
\cite{Zhang2015_mag,Lado2017,Webster2018,Richter2018,Lee2020,Bacaksiz2021,Ke2021,Cenker2021}
means that we must treat spins and atomic displacements on the same footing. We
develop a minimal model of this kind,
and show that the
adiabatic velocity-force approach can greatly overestimate the frequency splitting
of the chiral modes.

We begin by reviewing the formalism of the velocity-force coupling from previous
works \cite{qin12_berry_curvat_phonon_hall_effec,
  PhysRevLett.123.255901,PhysRevB.105.064303,
  mead79_deter_born_oppen_nuclear_motion,Bistoni2021}, which we will refer to  as
the ``Mead-Truhlar'' (MT) approach. In order to simplify the discussion, we
assume a finite system, generalizing to an infinite crystal below. The starting
point of the derivation is the Born-Oppenheimer approximation, where the system
wavefunction is factored into nuclear and electronic parts such that the
ground-state electronic wavefunction $\vert \psi(\vec{R}) \rangle$
depends parametrically on the nuclear
coordinates $\vec{R}$ \cite{BornHuang}. Once the electronic degrees of freedom
are integrated out, the effective Hamiltonian for the nuclear wavefunction
becomes \cite{mead79_deter_born_oppen_nuclear_motion}
\begin{equation}
\label{eq:Heff}
  H_{\mathrm{eff}} = \sum_{i\alpha}\frac{(p_{i\alpha}- \hbar A_{i\alpha}(\vec{R}))^{2}}{2 m_{i}} + V_{\mathrm{eff}}(\vec{R}),
\end{equation}
where Roman indices (here $i$) run over nuclei, Greek indices (here $\alpha$)
run over Cartesian directions, $p_{i\alpha}$ is the momentum operator for
nucleus $i$ along direction $\alpha$, and $m_i$ is the mass of nucleus $i$.
Using the notation $\partial_{i\alpha}=\partial/\partial R_{i\alpha}$,
$A_{i\alpha}(\vec{R}) = i \braket{\psi(\vec{R})\mid\partial_{i\alpha}\psi(\vec{R})}$
is a nuclear Berry potential, and
$V_{\mathrm{eff}}(\vec{R}) = \epsilon(\vec{R}) + \sum_{i\alpha}\frac{\hbar^{2}}{2m_{i}}\bigl(\braket{\partial_{i\alpha}\psi(\vec{R})\mid\partial_{i\alpha}\psi(\vec{R})} - A_{i\alpha}(\vec{R})^{2}\bigr)$
is an effective scalar potential, where $\epsilon(\vec{R})$ is the
ground-state energy for a given fixed nuclear configuration. As first pointed
out by Mead and Truhlar \cite{mead79_deter_born_oppen_nuclear_motion}, the
nuclear Berry potential $A_{i\alpha}$ cannot always be made to vanish by
changing the gauge of $\ket{\psi(\vec{R})}$ via the choice of an
$\vec{R}$-dependent phase factor.

The resulting vibrational modes are then found by solving the equations of
motion which, to harmonic order in nuclear displacements, are given by
\cite{qin12_berry_curvat_phonon_hall_effec,
  mead79_deter_born_oppen_nuclear_motion,
  PhysRevLett.123.255901,PhysRevB.105.064303}
\begin{equation}
  \label{eq:FV_geneig}
    \omega_{n}^{2}\,\mathbf{M}\,\eta_{n} = (\mathbf{C} + i\omega_{n}\mathbf{G})\eta_{n} .
\end{equation}
Here $\mathbf{M}$ is a diagonal nuclear mass matrix
$M_{i\alpha,j\beta} = m_{i}\delta_{i,j}\delta_{\alpha,\beta}$, $\omega_{n}$ is
the frequency of mode $n$, and $\eta_{n}(\tau\alpha)$ is the component of the
eigendisplacement of nucleus $\tau$ along direction $\alpha$ normalized so that $\eta_n^\dag\mathbf{M}\eta_m=\delta_{nm}$.
$\mathbf{G}$ is the ``velocity-force matrix,'' whose elements
$G_{i\alpha, j\beta}$ relate the force on nucleus $i$ along direction $\alpha$
to the velocity of nucleus $j$ along direction $\beta$,
and is expressed as
\begin{equation}
  \label{eq:force_velocity}
  G_{i\alpha,j\beta} =
  -2\hbar\mathrm{Im}\braket{\partial_{i\alpha}\psi(\vec{R})\mid\partial_{j\beta}\psi(\vec{R})},
\end{equation}
which is just $\hbar$ times the nuclear Berry curvature.
The matrix $\mathbf{C}$ is the IFC matrix, which we define to be
$C_{i\alpha,j\beta} = 
\partial_{i\alpha}\partial_{j\beta}\epsilon(\mathbf{R})$. Note that
following Eq.~(\ref{eq:Heff}) one could alternatively define $\mathbf{C}$ in
terms of the Hessian of $V_{\text{eff}}$, which includes additional terms compared to the conventional IFC.
However,
the additional terms are higher order in the inverse nuclear mass and do not
involve breaking of time-reversal symmetry, so that we neglect them in this work.

This formalism can be extended to the calculation of phonons in infinite
crystals within DFT. We restrict ourselves to phonons at $\Gamma$, the Brillouin
zone center, and work from a discrete set of DFT calculations on
primitive-cell structures in which each sublattice displacement in turn is
considered. The \textbf{C} matrix is constructed in the usual way from finite differences of the
forces, while \textbf{G} is built from Berry phases computed over triangular configuration
paths involving a given pair of sublattice-displaced structures and the
undistorted structure. Berry phases are computed on a per-unit-cell basis for
the Bloch manifold at each wave vector $\vec{k}$ and then averaged over the Brillouin zone
to get the desired \textbf{G} matrix elements. The details are given in
Sec.~S1 
of the supplemental material (SM) \cite{SM}.

We perform calculations on \cri{} in the ferromagnetic
ground state using the {\sc
  vasp} code
\cite{Kresse1993,Kresse1996,Kresse1999},
the local density approximation exchange-correlation functional
\cite{perdew1981}, and projector-augmented wave potentials \cite{Blochl1994}.
Semicore ($3s$, $3p$) and ($5s$, $5p$) states are included in the valence for
Cr and I respectively. A $5\times 5\times 5$ Monkhorst-Pack grid
\cite{PhysRevB.13.5188} is used to sample the Brillouin zone, and the energy
cutoff for the plane-wave basis set is 520 eV. Spin-orbit coupling, which is
essential to the physics described here, is included in all calculations.

\begin{figure}
  \includegraphics[width=\linewidth]{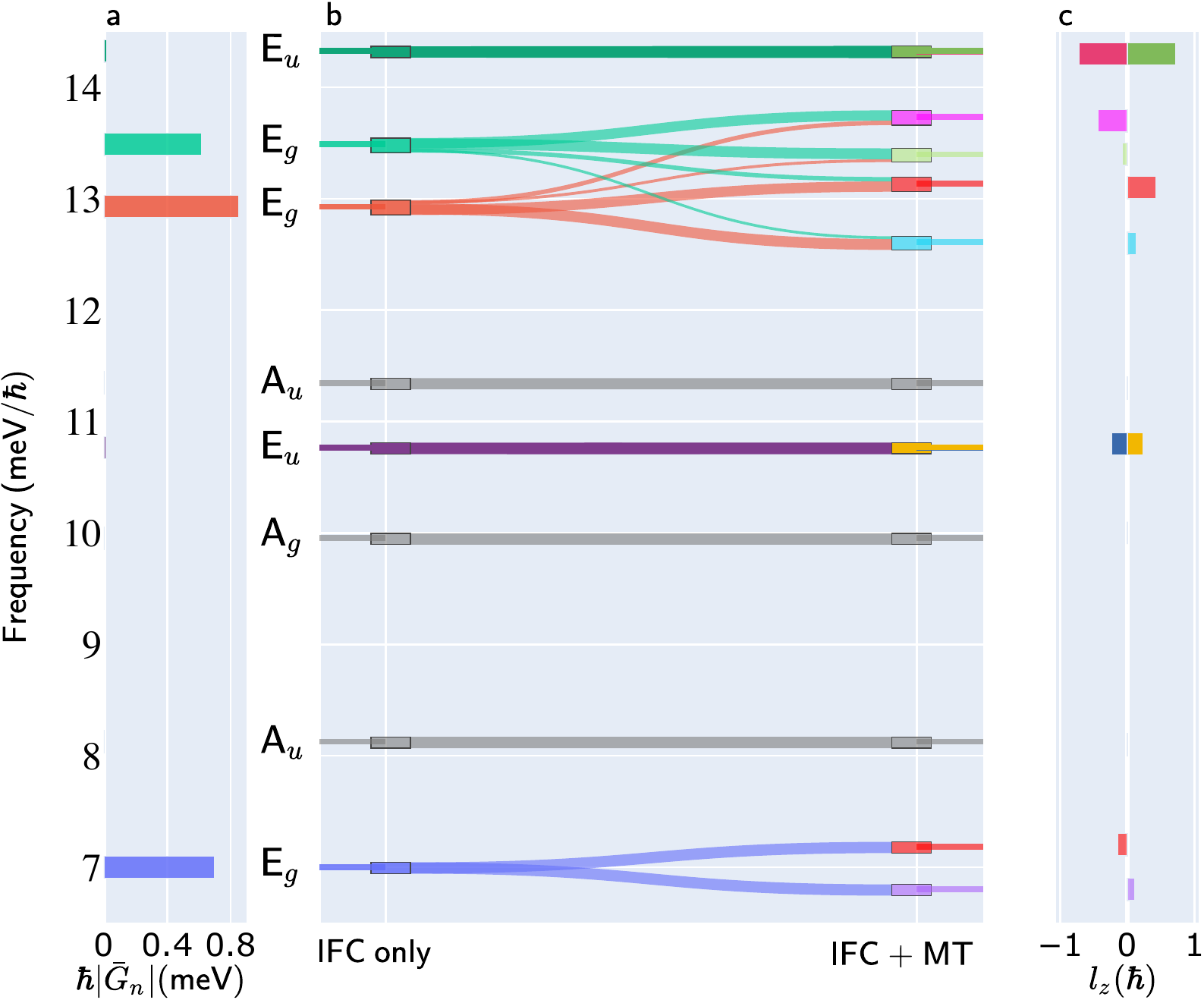}
  \caption{
    For selected zone-center phonon modes of bulk ferromagnetic CrI$_3$:
    (a)
  norm of the row of the velocity-force matrix $\bar{\mathbf{G}}$ relevant to each phonon mode
    (b) phonon
    frequencies (labeled by irreducible representation) determined from just the
    interatomic force-constant matrix on the left of the panel (``IFC only''), and including
    the
    Mead-Truhlar correction on the right side (``IFC +
    MT''), illustrating the frequency
    splitting of degenerate modes; and (c) angular momentum values of each mode
    after the velocity force matrix contributions have been included.
    In (b) the thickness of the connection between modes corresponds
    to the magnitude of the overlap between their respective eigenvectors (connecting curves are arbitrary).
  }
  \label{fig:splitting}
\end{figure}

As mentioned above, the conventional calculation of phonons neglects
\textbf{G} in Eq.~(\ref{eq:FV_geneig}), resulting in an equation of motion
with TRS ($\tilde{\omega}_{n}^{2}\,\mathbf{M}\,\tilde{\eta}_{n} = \mathbf{C}\tilde{\eta}_{n}$).
The resulting zone-center phonon modes with frequencies $\tilde{\omega}_{n}$
can be represented with real eigendisplacements $\tilde{\eta}_{n}$. In \cri{}
the representation of these modes at the zone center can be decomposed into the
real irreducible representations (irreps) of $\bar{3}$ as
$4A_g \oplus 4A_u \oplus 4E_g \oplus 4E_u$. The frequencies
$\tilde{\omega}_{n}$ range up to 32 meV (see the SM \cite{SM}
Table~SIII
). 
We can consider the velocity-force coupling between phonon modes obtained from the IFC alone by defining $\bar{G}_{nm}=\tilde{\eta_{n}}^{\dag}\mathbf{G}\tilde{\eta}_{m}$; $\bar{\mathbf{G}}$
is block diagonal, with each block corresponding to a real irrep. Once
\textbf{G} is included in Eq.~(\ref{eq:FV_geneig}), TRS is broken and the
twofold-degenerate $E_u$ and $E_g$ modes split, as the corresponding
irreps further break up into complex one-dimensional representations.

In Fig.~\ref{fig:splitting}(b), we show a selection of zone-center phonon
frequencies of bulk ferromagnetic CrI$_{3}$ (see the SM \cite{SM}
Table~SIII 
for a complete list of frequencies). On the left side we
 plot $\tilde{\omega}_n$, i.e., the frequencies neglecting the velocity-force
contribution, and on the
  right the frequencies including the velocity-force contribution via the MT approach (i.e.,
  $\omega_n$). In Fig.~\ref{fig:splitting}(a) we
plot the sum of the magnitudes of the velocity-force coupling terms relevant to
each phonon eigenvector, as a measure of the strength of the coupling. We see
that the $E_g$ modes have the
strongest coupling. We shall see why shortly.

The splitting of each of the two-fold degeneracies shown in
Fig.~\ref{fig:splitting}(b) results in chiral phonons with well-defined angular
momentum in the $z$ direction (due to the symmetry of \cri{}). For mode $n$, the
angular momentum is found via
$l_{z}^{(n)} = 2 \hbar \sum_{\tau} m_\tau \mathrm{Im}[\eta_{n}^{*}(\tau x)\eta_{n}(\tau y)]$\cite{Zhang2014, McLellan_1988}, where $\tau$ runs over the atomic sublattices and $m_\tau$ is the mass of nucleus $\tau$. If we
neglect \textbf{G} in Eq.~(\ref{eq:FV_geneig}), then the angular momentum of
modes in a degenerate subspace spanned by ($\tilde{\eta}_1, \tilde{\eta}_2$) depend on the basis we choose. Clearly, $l_z$
vanishes in the real basis $\tilde{\eta}_n$ introduced above, while a complex
``circularly polarized'' combination of degenerate modes of the form
$\eta'_{\pm}=\tilde{\eta}_{1} \pm i\tilde{\eta}_{2}$ will have equal and
opposite $l_{z}$ (since we are at the $\Gamma$ point
\cite{coh19_class_mater_with_phonon_angul}), with the magnitude determined by
the mode displacement patterns.
Neglecting terms in the $\bar{\mathbf{G}}$ matrix that
mix different degenerate subspaces, the chiral modes after splitting will have
exactly the eigendisplacements $\eta'_{\pm}$.
Fig 1 (c) shows the angular momentum ($l_{z}$) for each mode after the
  inclusion of the \textbf{G} matrix. The fact that some of the split modes do
not have equal and opposite $l_z$ indicates mixing between the degenerate
subspaces in $\bar{\mathbf{G}}$. See Sec.~S4 
of the SM for a full
analysis of the angular momentum of the chiral modes\cite{SM}.

We now analyze the mechanisms responsible for the velocity-force coupling in
CrI$_3$. In this material, the magnetic moments reside on the Cr atoms and are
initially oriented out of plane along $z$. We note that the modes with large
$\bar{\mathbf{G}}$ matrix elements in Fig.~\ref{fig:splitting} are mostly those
involving displacements of the I sublattices, which carry strong spin-orbit
coupling. Under such displacements, it is natural that the Cr moments may cant,
reflecting a local change in magnetic easy axis. This canting will result in a
spin Berry curvature, and thus a contribution to \textbf{G}.

In fact, we find
that for \cri{}, the spin Berry curvature is the dominant mechanism of
velocity-force coupling.
We demonstrate this by calculating the matrix elements of \textbf{G} under the
assumption that only the spin-Berry-curvature mechanism is present. 
More specifically, we first approximate Eq.~(\ref{eq:force_velocity}) as 
   $\bar{G}_{mn} \simeq B_{Ia,m} G_{Ia,Jb} B_{Jb,n}$, 
   where $G_{Ia,Jb}=-2\hbar \,\textrm{Im}\langle \partial_{Ia} \psi 
   \vert \partial_{Jb}\psi \rangle$ is the Berry curvature of the 
   wavefunctions in the parameter space spanned by the Cr spins; 
$B_{Ia,n}=\partial s_{Ia}/\partial \tilde{u}_{n}$ is a ``spin canting matrix''
describing the static
change in the equilibrium spin unit vector on magnetic Cr site $I$ in direction
$a$ resulting from phonon perturbation $n$; and $\tilde{u}_n$ is the amplitude of mode $n$ such that the set of atomic displacements are given by $\tilde{u}_n \tilde{\eta}_n$
\footnote{Here $a$ runs only over
  $x,y$ directions corresponding to the two possible tilt angles. We have
  assumed the total magnitude of the spin is unchanged and that tilt angles are
  small so that $S_{Ix}=S\sin(\theta_{Ix})\approx S s_{Ix}$}.
Under the assumption that
the spin Berry curvature dominates,
we can further approximate 
$G_{Ia,Jb}
= -S\,\delta_{IJ} \epsilon_{ab}$,
where spin $S=3\hbar/2$ for Cr and $\epsilon_{ab}$ is the antisymmetric tensor;
then
\begin{equation}
\label{eq:G_spin}
\bar{G}_{nm} = -S \sum_{Iab} \epsilon_{ab}\,B_{Ia,n}\,B_{Ib,m}.
\end{equation}

A comparison between this spin-Berry approximation and the full Berry-phase
calculation 
of the $\bf G$ matrix is presented in Table~SII 
of the
SM \cite{SM}.
For some modes, the
spin-Berry approximation accounts for only
$\sim60$\% of the full velocity-force contribution; this means that other
contributions, e.g., ``phonon-only Berry curvature'' (the Berry curvature from atomic
displacements at fixed spin) are significant.
However, these modes exhibit
relatively small splittings from the full velocity-force contribution, so the
errors are not significant in absolute terms.
For the two $E_g$ modes
that exhibit the largest splitting in the MT calculation
[green, orange, and blue curves in Fig.~\ref{fig:splitting}(b)], the errors arising from
this approximation are less than one percent.

This is a remarkable result, and is one of the main findings of the present
work. By adopting the spin-Berry approximation, one can bypass the cumbersome
computation of Berry phases for paths connecting distorted structures. Instead,
all we need to take from the DFT calculations is the information about the spin
canting in response to phonon distortions. In particular, it is now clear why
the $\mathbf{G}$ tensor elements are so much smaller for the E$_u$ modes; these
are the ones that couple to the optical magnons, whose much larger stiffness
strongly suppresses the spin canting.

A critical implication of the fact that the velocity-force coupling in \cri{}
results from spin canting is that the assumption underlying the MT approach of
Eqs.~(\ref{eq:Heff}-\ref{eq:force_velocity}), namely, that all electronic
dynamics are fast compared to that of the phonons, is clearly unfounded. This is
because the relevant time scale for spin dynamics is that of the magnon
frequencies in the system. The experimentally measured zone-center magnons of
\cri{} have frequencies of 0.3 meV (i.e., the magnetocrystalline anisotropy) for
the acoustic branch, and 17 meV for the optical branch \cite{Cenker2021}, while
the relevant phonons with the largest velocity-force coupling have frequencies
in the range of 6-14 meV [see Fig.~\ref{fig:splitting}(b)].

Thus, an appropriate description of the low-energy dynamics must treat spins and
phonons in this system on the same footing. To illustrate how this can be done,
we focus on a single pair of $E_g$ or $E_u$ modes, which couple respectively
either to an effective acoustic spin unit vector ${\bf s}=({\bf s}_1+{\bf s}_2)/\sqrt{2}$ or
its optical counterpart ${\bf s}=({\bf s}_1-{\bf s}_2)/\sqrt{2}$.
Denoting the phonon mode
amplitudes and momenta as $(x,y)$ and $(p_{x},p_{y})$,
the coupled spin-phonon Hamiltonian takes the form
\begin{equation}
  \label{eq:sp_model}
  \begin{split}
  H &= \frac{1}{2}(p_{x}^{2} + p_{y}^{2}) + \frac{1}{2}\tilde{\omega}^{2}(x^{2} + y^{2}) \\
    &+ \frac{1}{2}\alpha(s_{x}^{2} + s_{y}^{2})
    +\gamma(x s_{x} + y s_{y}) .
\end{split}
\end{equation}
Here $\tilde{\omega}$ is the bare phonon frequency,
 $\alpha = \partial^2 E/\partial^{2} s_x$
is the spin anisotropy energy, and
$\gamma = \partial^2 E/\partial x \partial s_x$
is the coupling between the spin and the pair of phonons (which have been chosen
such that the $xs_y$ and $ys_x$ terms vanish). Note that the unperturbed magnon frequency is
related to the anisotropy by $\omega_{\rm m}=\alpha/S$, where $S=3\hbar/2$ is the Cr
spin, and that the $\bf B$ matrix introduced above reduces in this minimal
model to $B=\gamma/\alpha$.
Going over to circularly polarized coordinates via
$x_{\pm} = (x \pm i y)/\sqrt{2}$ and
$s_{\pm} = (s_{x} \pm i s_{y})/\sqrt{2}$,
the equations of motion become
\begin{equation}
\label{eq:coup_eom}
\begin{split}
    (\tilde{\omega}^2-\omega^2)x_\pm &=-\gamma s_\pm, \\
(\pm\omega_{\rm m}-\omega)s_\pm &=\mp S^{-1}\gamma x_\pm,
\end{split}
\end{equation}
which are easily solved numerically.

\begin{table}
\begin{ruledtabular}
\begin{tabular}{lc|cc}
 & & \multicolumn{2}{c}{splitting (meV)}    \\
  irrep & $\hbar\tilde{\omega}$ (meV) &
                                        MT & SP  \\ \hline
 E$_g$      &    6.9999 &         0.3820 &      0.0007 \\
            &   12.9287 &         0.5270 &      0.0003 \\
            &   13.4876 &         0.3368 &      0.0001 \\
            &   29.8521 &         0.0244 &      $3\times 10^{-6}$ \\
\hline
 E$_u$      &   10.7667 &         0.0043 &      0.0046 \\
            &   14.3259 &         0.0090 &      0.0311 \\
            &   27.8168 &         0.0349 &      0.0118 \\
\end{tabular}
\end{ruledtabular}
  \caption{\label{tab:split} 
  Frequency splitting of $E_{u}$ and $E_{g}$ zone-center
    phonon modes in \cri{}.
      Modes are labeled by their
    symmetry and frequency determined only from the interatomic force constants
    ($\hbar\tilde{\omega}$). MT (``Mead-Truhlar'')
    refers to frequency splittings obtained by
    solving the equation of motion in Eq.~(\ref{eq:FV_geneig}), and SP (``spin-phonon'')
    corresponds to solving the coupled equations of motion in
    Eq.~(\ref{eq:coup_eom}). 
    }
\end{table}

In Table~\ref{tab:split}, we compare the frequency splittings of
doubly-degenerate modes determined by the adiabatic
Mead-Truhlar (MT) approach
and spin-phonon (SP) model.
The last column in Table~\ref{tab:split} presents the results for the
spin-phonon model; the $E_g$ modes couple to the acoustic magnon branch and the
$E_u$ modes to the optical branch (we use experimental \cite{Cenker2021} magnon frequencies of 0.3
and 17 meV, respectively) \footnote{A more general description
  of spin-phonon coupling, treating all relevant nuclear and spin degrees of
  freedom simultaneously, will be presented in a forthcoming communication.}.
A more detailed comparison of
Mead-Truhlar approach and spin-phonon model is
presented in Sec.~S2 
of the SM\cite{SM}.

\begin{figure}[ht]
  \centering
  \includegraphics[width=\linewidth]{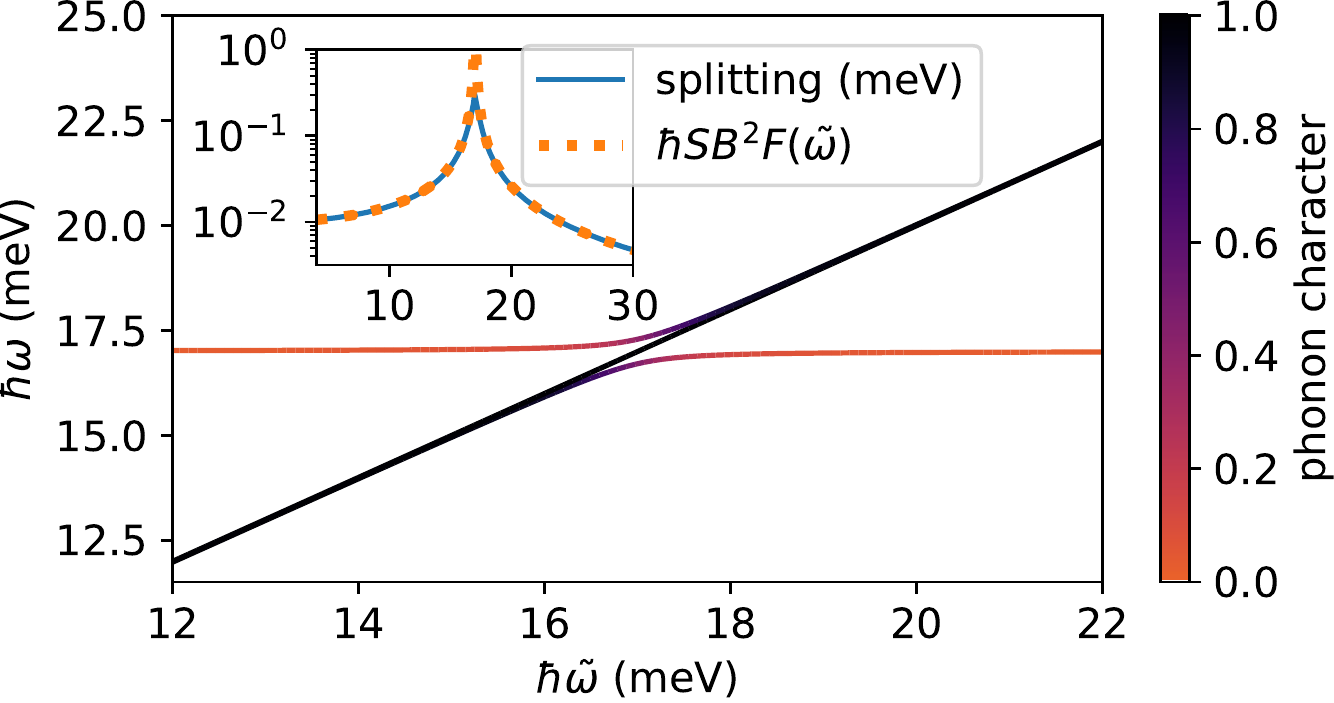}
  \caption{
    Frequencies of interacting magnons and phonons as given by
    Eq.~(\ref{eq:coup_eom}) as a function of uncoupled phonon frequency ($\tilde{\omega}$) for
$\omega_{\rm m}=17$meV$/\hbar$ and
$\gamma=2$  meV$^{3/2} / \hbar$
which corresponds to a force velocity matrix element of $\bar{G}=0.01 \textrm{meV}/\hbar$.
The color of the curve
indicates the magnitude of the phonon component of the mode eigenvector. The inset
shows the splitting of the modes as well as the heuristic for splitting away
from resonance, $\hbar SB^2 F(\tilde{\omega})$, where $F=|1 - (\tilde{\omega}/\omega_{\rm m})^2|^{-1}$.
\label{fig:model_v_phonon}}
\end{figure}

In order to understand the
general features of the spin-phonon mixing embodied in Eq.~(6),
we plot in Fig.~\ref{fig:model_v_phonon} the solutions
of Eq.~(\ref{eq:coup_eom}) as a function of phonon frequency $\tilde{\omega}$, with
the magnon frequency set to $\omega_{\rm m}=17$\,meV$/\hbar$ (i.e., the optical branch
\cite{Cenker2021}), and $\gamma$ fixed at
2\,meV$^{3/2}$/$\hbar$,
a typical value for the $E_u$ modes in \cri{} (since those are the phonons
that couple to the optical magnon).
The solid blue curve in the inset shows the splitting of the modes with dominant
phonon character also as a function of $\tilde{\omega}$. Outside of the small
``resonant'' regime $\tilde{\omega}\simeq\omega_{\rm m}$, where
significant magnon-phonon hybridization occurs, the mode splitting is well
described by $SB^2F$ (see orange dashed curve in inset of
Fig.~\ref{fig:model_v_phonon}), where $SB^2$ is the relevant force-velocity term
in the adiabatic spin-Berry approximation [i.e., Eq.~(\ref{eq:G_spin})] and
$F= |1 - (\tilde{\omega} / \omega_{\rm m})^2|^{-1}$. At small $\tilde{\omega}$,
the magnon can be treated as a high-energy degree of freedom which renormalizes
the phonons, and the splitting from the MT approach is recovered. Increasing
$\tilde{\omega}$ toward $\omega_{\rm m}$ enhances the splitting of modes over
the value at the adiabatic (MT) limit, peaking at the point where the magnon and
phonon frequencies coincide and the modes have maximum hybridization. Above
$\omega_{\rm m}$, the splitting decreases from its peak value, but is still
enhanced over that determined by the MT approach for a range of $\tilde{\omega}$. At
$\tilde{\omega} \gg \omega_{\rm m}$ the splitting is suppressed, and vanishes as
$\tilde{\omega}\rightarrow\infty$.

This behavior is reflected in the splittings given in Table~\ref{tab:split}.
Two of the three $E_u$ modes which couple to the optical magnon have frequencies
below $\omega_{\rm m}=17$\,meV$/\hbar$, but still in ranges where the splitting is enhanced above
the adiabatic MT limit. The physical interpretation here is that the
phonons are becoming hybridized with the magnon instead of just having their
frequencies renormalized by it. However, since the splittings in the adiabatic
limit (and thus the corresponding \textbf{G} matrix elements) are proportional to
$\omega_{\rm m}^{-2}$, the splittings remain quite small. The largest frequency $E_u$ mode
has $\omega_{\rm m} \ll \tilde{\omega}$,
so that the splitting is
reduced compared to values obtained from the velocity-force approach.

The $E_g$ modes couple to the acoustic magnon ($\omega_{\rm m}=0.3$ meV$/\hbar$), so for all
$E_g$ modes $\tilde{\omega} \gg \omega_{\rm m}$, which is precisely the opposite of the limit in
which the adiabatic velocity-force theory is applicable. This results in
the drastic reduction ($F \ll 1$) in the splitting of the $E_g$ modes
in Table~\ref{tab:split} compared to the
MT description in
Eq.~(\ref{eq:FV_geneig}). The physical
interpretation of this regime is that the Cr spins cannot keep up with the
phonons, and thus the area swept out from the spin canting is
greatly reduced compared to the assumption that they follow the nuclear motion
adiabatically.

Clearly, these results have significant implications for experimental measurements of chiral phonons in \cri{}. Optical techniques, possibly utilizing circular polarization, constitute a powerful tool for studying such properties \cite{Nafie1976,Zhu2018,Laiho1975,Du2019,Yin2021}. The strongly suppressed frequency splitting (SP column of Table~\ref{tab:split}) of the Raman-active $E_g$ modes is likely to be difficult to detect. This is consistent with recent work on CrBr$_3$, a related system, which found signatures of the chiral phonons but did not report a splitting of these modes \cite{Yin2021}. The larger, though still quite modest, splitting of the $E_u$ modes could in principle be measured by peak shifts in infrared absorption \cite{Tomarchio2021}, while direct detection of chirality would require circularly polarized infrared spectroscopy as in Refs.~\cite{Nafie1976, Laiho1975}.

More generally, our results also have several implications for finding other systems with a large
splitting of chiral phonon modes at $\Gamma$. In materials like \cri{}, where
the spin-Berry mechanism is responsible for the majority of the velocity-force
coupling, it is most promising to look for (or engineer via, e.g., strain or
magnetic field) cases where the relevant phonon and magnon frequencies coincide.
This avoids the suppression of the splitting in the
$\tilde{\omega} \gg \omega_{\rm m}$ regime, and the small spin canting likely
when $\omega_{\rm m} \gg \tilde{\omega}$.
Systems with lower-frequency optical magnons that maintain similar or larger
spin-phonon coupling $\gamma$ are also strong candidates for observing larger
effects.
In any case, we can see from Table~\ref{tab:split} that
correctly accounting for the relative dynamics of spins versus phonons is
necessary to avoid significantly overestimating the splitting of certain chiral
modes.

In conclusion, we have developed and implemented a first-principles
  methodology for capturing time-reversal-symmetry-broken lattice dynamics in
  magnetic solids, and applied it to the case of bulk ferromagnetic \cri{}. We
  show that in this system, the previously-made assumption of fast electron
  dynamics compared to the lattice breaks down, since the relevant coupling is
  between Cr spins and atomic displacements. With a minimal model, we
  demonstrate that spins and phonons must be treated on the same footing to
  avoid large qualitative
  errors in the splitting of chiral modes. 

\acknowledgements

CED, DV, and SC acknowledge support from the National Science Foundation under Grants
DMR-1918455, DMR-1954856, and DMR-1848074 respectively. The Flatiron Institute is a
division of the Simons Foundation.
M.S. acknowledges the support of Ministerio de Ciencia y Innovaci\'on (MICINN-Spain) through
 Grants No. PID2019-108573GB-C22 and CEX2019-000917-S;
 of Generalitat de Catalunya (Grant No. 2017 SGR1506);
 and of the European Research Council (ERC) through Grant No. 724529.

\bibliography{forcevel.bib,CED_bib.bib}{}
\end{document}


\title{Supplemental Material: Frequency splitting of chiral phonons from broken time reversal symmetry in CrI$_3$}
\author{John Bonini}
\affiliation{Center for Computational Quantum Physics, Flatiron Institute, 162 5th Avenue, New York, New York 10010,
USA}
\author{Shang Ren}
\affiliation{Department of Physics and Astronomy, Rutgers University, Piscataway, New Jersey 08845-0849, USA}
\author{David Vanderbilt}
\affiliation{Department of Physics and Astronomy, Rutgers University, Piscataway, New Jersey 08845-0849, USA}
\author{Massimiliano Stengel}
\affiliation{Institut de Ci\`{e}ncia de Materials de Barcelona (ICMAB-CSIC), Campus UAB, 08193 Bellaterra, Spain}
\affiliation{ICREA-Instituci\'{o} Catalana de Recerca i Estudis Avan\c{c}ats, 08010 Barcelona, Spain}
\author{Cyrus E. Dreyer}
\affiliation{Department of Physics and Astronomy, Stony Brook University, Stony Brook, New York, 11794-3800,
  USA}
\affiliation{Center for Computational Quantum Physics, Flatiron Institute, 162 5th Avenue, New York, New York 10010,
USA}
\author{Sinisa Coh}
\affiliation{Materials Science and Mechanical Engineering, University of California Riverside, CA 92521, USA}

\date{\today}
\maketitle

\section{Finite difference approach for calculating Berry curvatures \label{sec:diff}}

Here we provide some details of our density-functional theory (DFT)
finite-displacement approach for calculating the interatomic force constant
(IFC) matrix \textbf{C}, and the adiabatic velocity-force coupling \textbf{G}. The IFC matrix is
calculated in the conventional way \cite{gonze1997,Baroni_2001} by displacing each
nucleus $i$ by distance $\delta\tau$\,=\,0.015\,\AA{} in each Cartesian direction,
and calculating the resulting forces.  The wave functions from these displaced
calculations are saved for the following steps.
The full velocity-force matrix elements [see Eq.~3 
in the main text] are calculated
by considering closed paths among these displaced configurations,
as discussed in Sec.~\ref{sec:full}, while the matrix elements
\textbf{G} under the spin-Berry approximation [see main text
Eq.~4
] are determined by calculating the canting of the Cr spins
from the same displacements, see Sec.~\ref{sec:spinberry}.

\subsection{Velocity-force coupling \label{sec:full}}
Stokes' theorem equates the integral of the Berry curvature over some
region in parameter space with the integral of the Berry potential along the
boundary of that region. We consider the Berry phase $\phi_{i\alpha,j\beta}$ 
corresponding to a state evolving adiabatically along a triangular path
from the ground state structure to displacement $\tau_{i\alpha}$, then to
displacement $\tau_{j\beta}$, and back to the ground state.  Assuming
$\delta\tau$ is small enough that the Berry curvature is constant over
the area of the path,
\begin{equation}
\label{eq:Omega}
  G_{i\alpha,j\beta} = \frac{2\hbar\phi_{i\alpha,j\beta}}{\mid\tau_{i\alpha} \wedge \tau_{j\beta}\mid},
\end{equation}
where $\tau_{i\alpha}$ is a displacement of nucleus $i$ in direction $\alpha$.
Since these displacements are chosen as orthogonal, the wedge product is simply
$\delta\tau^2$.
For a finite system, letting $\psi(\tau_{i\alpha})$ be the ground state electronic
wave function for the displaced structure, we compute the three-point
discrete Berry phase
\begin{equation}
  \phi_{i\alpha,j\beta} = -\mathrm{Im}\ln\left[\braket{\psi(0) \mid \psi(\tau_{i\alpha})}\braket{\psi(\tau_{i\alpha}) \mid \psi(\tau_{j\beta})}\braket{\psi(\tau_{j\beta}) \mid \psi(0)}\right],
  \label{eq:phi}
\end{equation}
where $\psi(0)$ indicates the ground state without displacements. For a solid
calculated within DFT, Eq.~(\ref{eq:phi}) holds for the single-band case, except
that we define $\psi_{\textbf{k}}(\tau_{i\alpha})$ to be the Kohn-Sham Bloch wavefunctions for
the displaced structures, and we must average over $k$ points in the first
Brillouin zone.  That is,
\begin{equation}
    \phi_{i\alpha,j\beta} =  -\frac{1}{N_{\textbf{k}}} \sum_{\textbf{k}} \mathrm{Im}\ln\left[\braket{\psi_{\textbf{k}}(0) \mid \psi_{\textbf{k}}(\tau_{i\alpha})}\braket{\psi_{\textbf{k}}(\tau_{i\alpha}) \mid \psi_{\textbf{k}}(\tau_{j\beta})}\braket{\psi_{\textbf{k}}(\tau_{j\beta}) \mid \psi_{\textbf{k}}(0)}\right] \,,
  \label{eq:phi-sum-k}
\end{equation}
where \textbf{k} is the Bloch wave vector and $N_{\textbf{k}}$ is number of $k$ points in the Brillouin zone. 
The generalization of the discrete Berry phase for multiple bands can be found
in Ref.~\onlinecite{vanderbilt_2018_3}, and is given by
\begin{equation}
    \phi_{i\alpha,j\beta} = -\frac{1}{N_{\textbf{k}}} \sum_{\textbf{k}} \mathrm{Im} \ln \mathrm{Det}[ M^k(0,\tau_{i\alpha}) M^k(\tau_{i\alpha},\tau_{j\beta}) M^k(\tau_{j\beta},0) ] \,,
\end{equation}
where the overlap matrices are defined as
\begin{equation}
    M_{mn}^k (\tau,\tau') = \ip{\psi_{mk}(\tau)}{\psi_{nk}(\tau')}
\end{equation}
with $m$ and $n$ being the band indices.

\subsection{Velocity-force coupling under spin-Berry approximation \label{sec:spinberry}}

Within the spin-Berry approximation of Eq.~4
, the matrix elements of
\textbf{G} are obtained from the response of the localized spins to atomic
displacements. The assumption is that for a given closed path along which
nuclear coordinates are changed, the Berry phase is given by the sum of the Berry
phases picked up by each local spin along that path. Each spin picks up a phase
equal minus to the total spin magnitude, $-S$ (in our case $S=3/2$), times the solid angle
on the Bloch sphere that is swept out along the path. For small canting angles
and with all spins pointing along $z$ in the ground state, \textbf{G} is then
constructed using Eq.~\ref{eq:Omega} with $\phi_{i\alpha,j\beta}$ now given by
\begin{equation}
  \phi_{i\alpha, j\beta} =  - \frac{S}{2} \sum_{I} (s_{Ix}(\tau_{i\alpha})s_{Iy}(\tau_{j\beta}) - s_{Iy}(\tau_{i\alpha})s_{Ix}(\tau_{j\beta}))\,.
\end{equation}
Equivalently, \textbf{G} could be constructed from the
derivatives of the spins with respect to atomic displacements as in
Eq.~4
\def\PL{\phantom{-}}

\begin{table}[h]
\begin{ruledtabular}
\centering
\begin{tabular}{cc|rrrr} 
 Irrep & $\hbar\tilde{\omega}$ (meV)& $s_{1x}$\PL\PL\PL & $s_{1y}$\PL\PL\PL & $s_{2x}$\PL\PL\PL & $s_{2y}$\PL\PL\PL  \\
 \hline
\multirow{8}{*}{$E_g$}&\multirow{2}{*}{6.9999}  & $-0.01047$  & $ \PL0.00416$  & $-0.01047$  & $ \PL0.00416$ \\
&  & $-0.00416$  & $-0.01047$  & $-0.00416$  & $-0.01047$ \\
 \cline{2-6}
&\multirow{2}{*}{12.9287}  & $-0.00253$  & $-0.01356$  & $-0.00253$  & $-0.01356$ \\
&  & $-0.01356$  & $ 0.00253$  & $-0.01356$  & $0.00253$ \\
 \cline{2-6}
&\multirow{2}{*}{13.4876} & $ 0.00521$  & $-0.00847$  & $ 0.00521$  & $-0.00847$ \\
&  & $-0.00847$  & $-0.00521$  & $-0.00847$  & $-0.00521$ \\ 
 \cline{2-6}
&\multirow{2}{*}{29.8521}  & $ 0.00028$  & $-0.00317$  & $ 0.00028$  & $-0.00317$ \\
&  & $ 0.00317$  & $ 0.00028$  & $ 0.00317$  & $ 0.00028$ \\
\hline
\multirow{6}{*}{$E_u$}&\multirow{2}{*}{10.7667}  & $ 0.00008$  & $-0.00095$  & $-0.00008$  & $ 0.00095$ \\
&  & $ 0.00095$  & $ 0.00008$  & $-0.00095$  & $-0.00008$ \\
\cline{2-6}
&\multirow{2}{*}{14.3259}  & $ 0.00101$  & $ 0.00142$  & $-0.00101$  & $-0.00142$ \\
&  & $ 0.00142$  & $-0.00101$  & $-0.00142$  & $ 0.00101$ \\
\cline{2-6}
&\multirow{2}{*}{27.8168}  & $ 0.00251$  & $-0.00057$  & $-0.00251$  & $ 0.00057$ \\
&  & $-0.00057$  & $-0.00251$  & $ 0.00057$  & $ 0.00251$ \\
\end{tabular}
\caption{Derivative of spin canting of Cr atoms 1 and 2 in Cartesian
  directions $x$ and $y$ with respect to amplitudes of $E_g$ and $E_u$ modes
  in ferromagnetic \cri. 
  The units for the spin derivatives with respect to mode amplitude are $\sqrt{\mathrm{meV}}/\hbar$. 
  Degenerate mode pairs are labeled
  by their unperturbed frequencies $\hbar\tilde{\omega}$; the two rows for each
  entry correspond to the two modes comprising the pair in a real basis.
}
\label{table:eg-spin}
\end{ruledtabular}
\end{table}

\section{Contribution of spin canting to adiabatic velocity-force coupling \label{sec:spin_cant}}

In Table~\ref{tab:spin_berry_compare} we compare the velocity-force coupling
computed under the Mead-Truhlar (MT) approach (i.e., assuming adiabatic
electron dynamics), using the \textbf{G} matrix obtained from the the full
wave function Berry curvature as in Eq.~3 
of the main
text (labeled ``W''), with the \textbf{G} matrix obtained from the spin-Berry
approximation of Eq.~4 
of the main text (labeled ``S'').
Modes are labeled by their irrep and frequency as determined only from the IFC matrix ($\tilde{\omega}$).
For each pair we show the magnitude of the matrix
element between the degenerate modes ($G_{ij}$) and frequency splitting
$\Delta \omega$. Deviations between matrix elements and frequency splitting
occur only when \textbf{G} induces mixing between the degenerate subspaces. We
see that indeed, the spin-Berry contribution accounts for the majority of
spin-phonon coupling. For the three lowest-frequency $E_g$ modes and the $E_{u}$
mode near 14 meV, the agreement is excellent, within 1\%. For the highest-frequency
$E_g$ mode and the other two $E_u$ modes, the spin-Berry contribution
has a discrepancy of 25-43\%. This indicates that other terms, such as the
``phonon-only Berry curvature'' resulting from atomic displacements at fixed spins, may
have a significant contribution. In any case, within the MT theory,
the splittings of the highest-frequency $E_g$ mode and all of the $E_u$ modes are
quite small, so the error in absolute terms is not large.

\begin{table}[h]
\begin{ruledtabular}
\begin{tabular}{cc|cc|cc}
  & &\multicolumn{2}{c|}{MT (W)}&\multicolumn{2}{c}{MT (S)}    \\
 Irrep   &   $\hbar\tilde{\omega}$ (meV) &   $\hbar\bar{G}_{ij}$ (meV) &   $\hbar \Delta \omega$ (meV)&   $\hbar\bar{G}_{ij}$ (meV)&   $\hbar \Delta \omega$ (meV) \\
\hline
 E$_g$      &  6.9999 & 0.3825 &  0.3820 & 0.3808 &  0.3803 \\
            & 12.9287 & 0.5685 &  0.5270 & 0.5712 &  0.5293 \\
            & 13.4876 & 0.2948 &  0.3368 & 0.2968 &  0.3391 \\
            & 29.8521 & 0.0243 &  0.0244 & 0.0305 &  0.0306 \\
\hline
 E$_u$      & 10.7667 & 0.0043 &  0.0043 & 0.0027 &  0.0027 \\
            & 14.3259 & 0.0090 &  0.0090 & 0.0091 &  0.0091 \\
            & 27.8168 & 0.0349 &  0.0349 & 0.0199 &  0.0199 \\

\end{tabular}
\caption{\label{tab:spin_berry_compare} Comparison of velocity-force matrix
  elements ($\bar{G}_{ij} = \eta_i^\dag \mathbf{G} \eta_j$) and frequency splitting ($\Delta \omega$) of $E_{u}$ and
  $E_{g}$ zone center phonon modes in \cri{} computed with Berry curvatures
  obtained from wavefunction overlaps (W) and with Berry curvatures obtained by
  the solid angle swept out by magnetic moments on the Cr sites (S). Modes are
  labelled by their irrep and frequency determined only from the interatomic
  force constants ($\hbar\tilde{\omega}$).}
    \end{ruledtabular}
\end{table}

\section{Frequencies of all modes \label{sec:freqs}}
In the spin-phonon model (SP) of Eq.~5 
the $\gamma$ parameter
is related to the magnitude of the $\bar{\mathbf{G}}$ matrix elements coupling degenerate modes
$i$ and $j$ by $\gamma = \omega_{m}\sqrt{S\bar{G}_{ij}}$. Here the natural choice
for the $\bar{G}_{ij}$ value from which to obtain $\gamma$ is that of the spin-Berry
approximation, as $\gamma$ is really defined in terms of spin-phonon coupling.
It is in general possible for a system to have a nonzero MT term even when evaluated
at fixed spin, as well as a distinct $\gamma$ coupling to spin degrees of freedom; this
will be explored in subsequent work. While the SP values presented here
incorporate nonadiabatic effects not present in the nuclear MT theory, they do not
include any coupling of modes outside the
degenerate subspace (that is, there is no mixing between modes with distinct
$\tilde{\omega}$).

In order to present the effects of these different levels of approximation
independently, we consider four possible \textbf{G} matrices for MT (labeled W, w,
S, and s) and two possible \textbf{G} matrices for SP (w and s). As above, W and
S correspond to \textbf{G} matrices obtained using Eq.~3 
and Eq.~4 
of the main text, respectively. Results labeled ``w'' and ``s'' correspond to
setting all couplings between nondegenerate $\tilde{\omega}$ equal to zero in
``W'' and ``S'' \textbf{G} matrices respectively.
In Table~\ref{tab:modes} we give the frequencies of all of the optical modes
(including the singly-degenerate ones), from the IFC matrix only
($\hbar\tilde{\omega}$), with adiabatic Mead-Truhlar (MT), and in the
spin-phonon model (SP), for each of these cases.

\begin{table}[b]
\begin{ruledtabular}
\setlength{\tabcolsep}{12pt}
\begin{tabular}{r|rrr|rrr}
  $\hbar\tilde{\omega}$ & MT (W) & MT (w) & SP (w) & MT (S) & MT (s) & SP (s) \\
\hline
  6.9999 &  6.8021 &  6.8113 &  6.9996 &  6.8030 &  6.8122 &  6.9996 \\
  6.9999 &  7.1841 &  7.1938 &  7.0003 &  7.1832 &  7.1929 &  7.0003 \\
  8.1265 &  8.1265 &  8.1265 &  8.1265 &  8.1265 &  8.1265 &  8.1265 \\
  9.9535 &  9.9535 &  9.9535 &  9.9535 &  9.9535 &  9.9535 &  9.9535 \\
 10.7667 & 10.7645 & 10.7645 & 10.7575 & 10.7653 & 10.7653 & 10.7608 \\
 10.7667 & 10.7688 & 10.7688 & 10.7646 & 10.7680 & 10.7680 & 10.7653 \\
 11.3471 & 11.3471 & 11.3471 & 11.3471 & 11.3471 & 11.3471 & 11.3471 \\
 12.9287 & 12.6100 & 12.6475 & 12.9285 & 12.6082 & 12.6462 & 12.9285 \\
 12.9287 & 13.1369 & 13.2161 & 12.9288 & 13.1375 & 13.2174 & 12.9288 \\
 13.4876 & 13.3989 & 13.3410 & 13.4875 & 13.3983 & 13.3400 & 13.4875 \\
 13.4876 & 13.7357 & 13.6358 & 13.4877 & 13.7374 & 13.6368 & 13.4877 \\
 14.3259 & 14.3214 & 14.3214 & 14.2924 & 14.3213 & 14.3213 & 14.2919 \\
 14.3259 & 14.3304 & 14.3304 & 14.3230 & 14.3305 & 14.3305 & 14.3230 \\
 16.5013 & 16.5013 & 16.5013 & 16.5013 & 16.5013 & 16.5013 & 16.5013 \\
 16.6023 & 16.6023 & 16.6023 & 16.6023 & 16.6023 & 16.6023 & 16.6023 \\
 26.5081 & 26.5081 & 26.5081 & 26.5081 & 26.5081 & 26.5081 & 26.5081 \\
 27.8168 & 27.7993 & 27.7993 & 27.8127 & 27.8068 & 27.8068 & 27.8145 \\
 27.8168 & 27.8342 & 27.8342 & 27.8335 & 27.8267 & 27.8267 & 27.8263 \\
 29.8521 & 29.8407 & 29.8400 & 29.8521 & 29.8376 & 29.8369 & 29.8521 \\
 29.8521 & 29.8650 & 29.8643 & 29.8521 & 29.8681 & 29.8674 & 29.8521 \\
 31.7361 & 31.7361 & 31.7361 & 31.7361 & 31.7361 & 31.7361 & 31.7361 \\

\end{tabular}
\caption{Frequencies in meV of optical phonon modes of ferromagnetic \cri{} at the
  zone-center computed with different levels of theory. $\hbar \tilde{\omega}$
  labels frequencies computed using only the force constant matrix. MT refers to
  frequencies computed within the adiabatic velocity-force, or Mead-Truhlar formalism. SP refers to
  frequencies computed in the spin-phonon model. Labels W, w, S, and s refer to
  different couplings used in each calculation as explained in
  Sec.~\ref{sec:freqs}. }

\label{tab:modes}
\end{ruledtabular}
\end{table}

\section{Angular momentum of chiral phonons \label{sec:ang_mom}}

\begin{figure}
  \includegraphics[width=0.45\linewidth]{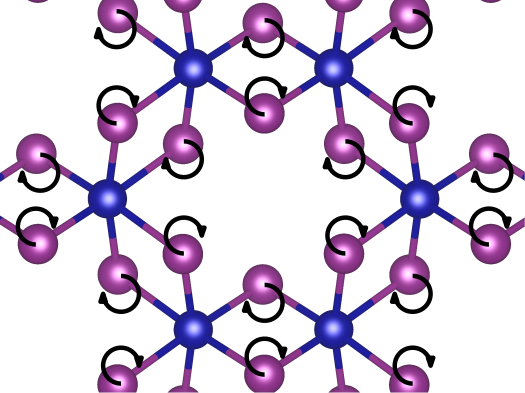}
  \hspace{9pt}
  \includegraphics[width=0.45\linewidth]{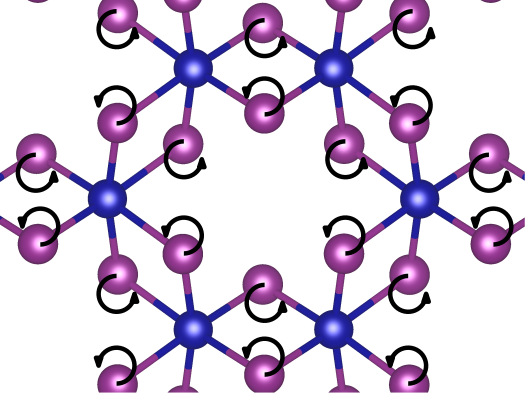}
  \caption{
    Schematic of chiral trajectories of selected zone-center phonon modes in a single layer of bulk 
    \cri{}. In the presence of time-reversal symmetry breaking, such phonons occur at
    distinct frequencies.}
  \label{fig:structures}
\end{figure}

In the main text we discussed the fact that including TRS breaking via \textbf{G} in the
equations of motion resulted in chiral phonons with well-defined angular
momenta. We now analyze in more detail the angular momentum of the phonon modes
shown in Fig.~1
(b) in the main text. From the symmetry of
\cri{}, specifically the three-fold rotation axis in the out-of-plane $z$
direction, the total angular momentum of a normal mode
can only have $z$ components. For mode $n$, the angular momentum is
$l_{z}^{(n)} = 2 \hbar \mathrm{Im}[\sum_{\tau}m_\tau\eta_{n}(\tau x)^{*}\eta_{n}(\tau y)]$, where
$\eta_{n}(i\alpha)$ is the eigendisplacement component for nucleus $i$
along direction $\alpha$ for the $n$th mode. Consider, for example,
the green $E_{g}$ mode in Fig.~1
(b) of the main text with
unperturbed frequency $\hbar\tilde{\omega}=13.5$ meV. The angular momentum of this mode
mode is due to the in-plane motion of iodine atoms. 
Choosing the two
degenerate modes $\tilde{\eta}_x$ and $\tilde{\eta}_y$ to involve real
displacements of I atoms moving in the $x$ and $y$ directions respectively,
neither of these modes has a non-zero angular momentum by itself. However,
the complex circularly polarized combination is
$\eta'_{\pm}=(\tilde{\eta}_{x} \pm i\tilde{\eta}_{y})/\sqrt{2}$, which corresponds to the
circular displacement patterns shown in Fig.~\ref{fig:structures}. 
In Fig.~1
(c) of the main text, we plot the angular momentum of the modes after
including the velocity-force coupling, with the color corresponding to the mode
on the right side of Fig.~1
(b). 
For the $E_{g}$ modes near 13 meV [green and orange in Fig.~1
(b)]
an analysis of the displacement patterns and angular momentum of the split modes is complicated by the fact that all four modes can mix [unlike the higher-frequency split mode and the gray $E_{u}$ mode above it in
Fig.~1
(b)].
Thus the angular momentum contributions seen in
Fig.~1
(c) are mixed between these modes.

A large velocity-force coupling does not necessarily result in a large phonon
angular momenta after splitting, as seen for the case the lowest two $E_{g}$ modes [blue
curves in Fig.~1
(b) of the main text]. The
relatively modest angular momentum of these modes [see
Fig.~1
(c) of the main text] is a result of the fact that the
motions of the real eigendisplacements before the inclusion of the \textbf{G}
perturbation involve \emph{out-of-plane} motions of I
atoms, not just in-plane motions like those depicted in Fig.~\ref{fig:structures}. This
is allowed by symmetry since individual atoms may have in-plane angular momentum
components $l_x$ and $l_y$, as long as these components sum to zero over all
atoms. The resulting $l_z$ is thus significantly reduced compared to the
green 13.5 meV $E_g$ mode. Note that even though the blue $E_g$ mode is quite
isolated in frequency compared to the other $E_g$ modes, the asymmetry of
$l_z$ of the modes after splitting indicates significant mixing with these other
modes.

Finally, we can see in Fig.~1
(c) of the main text that higher
frequency modes tend to have higher angular momentum. This is likely due to the
fact that the in-plane vibrations in \cri{} tend to have higher frequency, and,
as pointed out above, only in-plane displacements contribute to $l_z$.

\bibliography{forcevel.bib,CED_bib.bib}{}